\documentclass[10pt,letterpaper,twocolumn]{article} 

\usepackage[usenames,dvipsnames]{color}
\usepackage{ol2}
\usepackage{amsmath,array,amssymb}
\usepackage{graphicx}
\usepackage{hyperref}
\usepackage{textcomp}
\usepackage{braket}
\usepackage{multirow}
\usepackage{indentfirst}
\usepackage[FIGTOPCAP,raggedright,nooneline,bf,footnotesize]{subfigure}
\usepackage{paralist}

\begin{document}
\twocolumn[ 
\title{ Down-conversion source of positively spectrally correlated and decorrelated photon pairs at telecom wavelength }
\author{Thomas Lutz,$^{1,2,*}$ Piotr Kolenderski,$^{1,3,*}$ and Thomas Jennewein$^{1}$}
\address{
$^1$Institute for Quantum Computing, University of Waterloo, 200 University Ave.~West, Waterloo, Ontario, CA
N2L 3G1
\\
$^2$Institut f\"ur Quantenmaterie, Universit\"at Ulm, 89069 Ulm, Germany \\
$^3$Institute of Physics, Nicolaus Copernicus University, Grudziadzka 5, 87-100 Toru{\'n}, Poland\\
$^*$Corresponding author: tlutz@uwaterloo.ca, kolenderski@fizyka.umk.pl
}

\begin{abstract}
The frequency correlation (or decorrelation) of photon pairs is of great importance in long-range quantum communications and photonic quantum computing. We experimentally characterize a spontaneous parametric down conversion (SPDC) source, based on a $\beta$-Barium Borate (BBO) crystal cut for type-II phase matching at $1550$ nm which emits photons with the positive or no spectral correlations. Our system employs a carefully designed detection method exploiting two InGaAs detectors.
\end{abstract}
] 

Many quantum information processing protocols benefit from photon pairs that are positively correlated or uncorrelated in their spectra. Optical quantum gates often require pure states, which can be created using spontaneous parametric down-conversion (SPDC) only if there are no correlations within the resulting photon pair. On the other hand, practical quantum communication using existing infrastructure suffers from noisy detectors, which limit the achievable range. This could be improved by utilizing photon pairs that are positively correlated. Spectral filtering \cite{Dragan2004,Uren2005,Uren2007,Osorio2008,Mosley2008,Kolenderski2009} makes it possible to minimize correlations, but positive correlations can be achieved only by careful source design. 
In a typical scenario, the SPDC process naturally leads to spectral anti-correlations \cite{Kim2005,Wasilewski2006,Poh2007,Poh2009,Soeller2010}. Positively spectrally correlated photons at 1500 nm were suggested in theory \cite{Kim2002} but were never shown in experiment. Our particular source allows for anti-, positive- or no-spectral correlations without the need for spectral filtering.

In the SPDC process, one photon of the pump converts into two daughter photons. Energy and momentum conservation relations, jointly described as phase matching,  and the properties of the pumping photons, govern the characteristics of the generated photons. The probability amplitude governing a photon pair emission in a given direction and at a given frequency can be described by a product of the pump spatio-temporal amplitude and the phase matching function \cite{Kim2002,Kolenderski2009}. The phase matching describes the properties of the nonlinear media and specifies the allowed emissions that can take place. 

Typically, the output photons are coupled into optical fibers. This corresponds to collecting the photons from a specific range of directions, defined by the fiber and the optics. From this point of view, coupling can be understood as an additional condition to phase matching. Therefore, one can introduce the effective phase matching function (EPMF) $\Theta(\omega_s,\omega_i)$ \cite{Kolenderski2009}, which fully describes the joint effect of the crystal and coupling into fibers. This in turn allows us to express the  wave function of a fiber-coupled photon pair $\psi(\omega_s,\omega_i)$ as an overlap of EPMF and the pump spectral amplitude $A(\omega_s+\omega_i)$:
\begin{equation}
\psi(\omega_s,\omega_i)=\Theta(\omega_s,\omega_i) A(\omega_s+\omega_i).
\label{eq:epmf}
\end{equation}
The pump amplitude in the case of continuous wave (CW) pumping can be modelled using a Dirac delta function $A(\omega_s+\omega_i)\propto \delta(\omega_p-\omega_s-\omega_i)$.  In this case, the joint spectrum corresponds directly to the diagonal slice of EPMF, which according to \eqref{eq:epmf} is given as
\begin{equation}
\psi(\omega_s,\omega_p-\omega_s)\propto \Theta(\omega_s,\omega_p-\omega_s).
\end{equation}
By tuning the pump frequencies $\omega_p$ and measuring the joint spectrum $|\psi(\omega_s,\omega_i)|^2$, one can measure the slices of EPMF which, together with the pump envelope, provide all necessary information about the SPDC photons. 

\begin{figure}
\centerline{\includegraphics[width=\columnwidth]{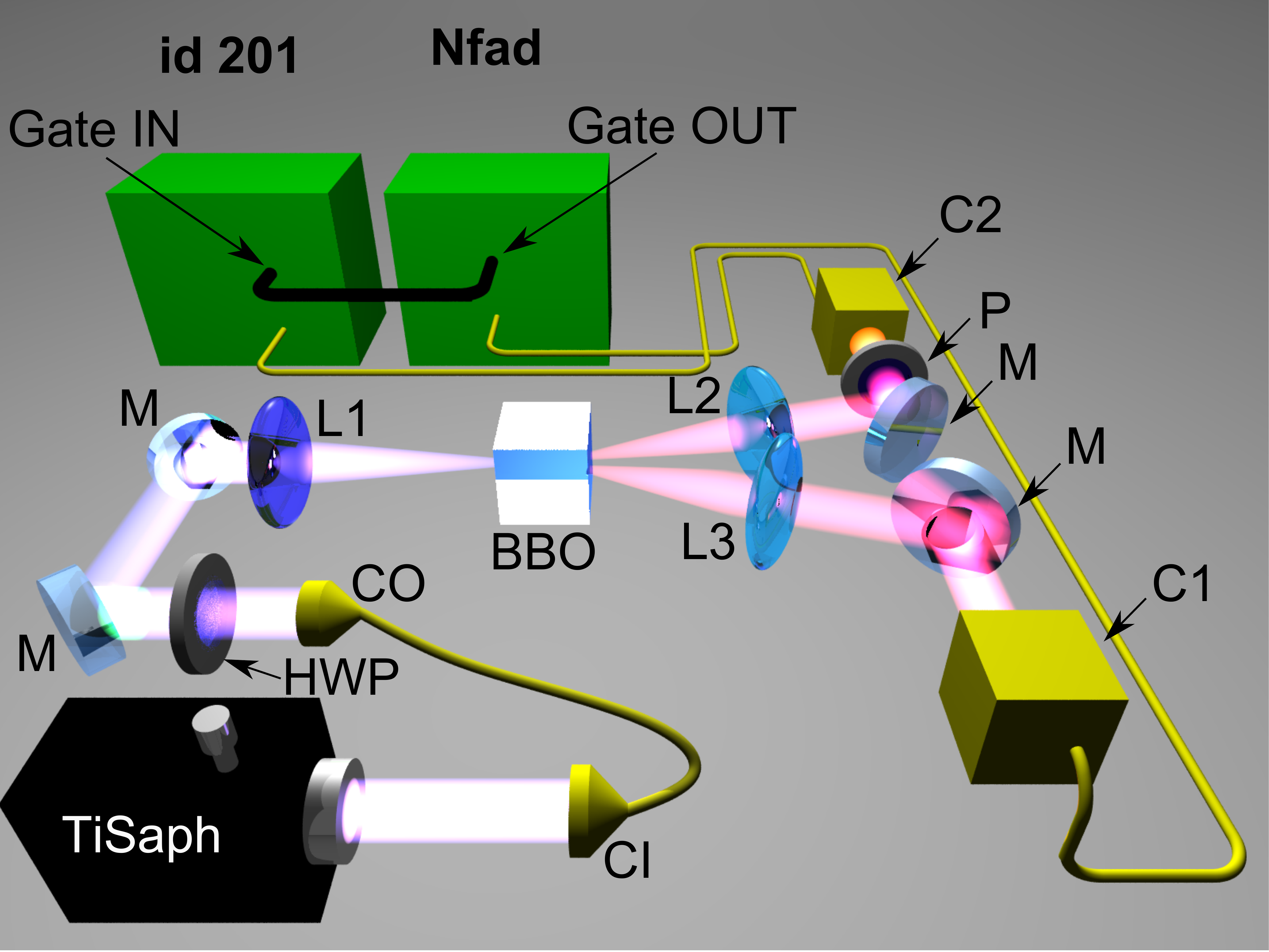}}
\caption{(Color online) Experimental setup. A tunable Ti:Sapphire laser is coupled into single mode fiber using a coupler (CI) with collimated output (CO). A half-wave plate (HWP) allows polarization correction. A lens (L1) focuses the pump, resulting in a beam radius of $150$ $\mu$m in the BBO crystal. Down-converted photons exit at an angle of $3$\textdegree\ and are collected by a system of plano-convex lenses (L2 and L3, focal length $150$ mm), and coupling aspheric lenses  (C1 and C2, focal length $15.4$ mm). The detection of one photon by an the NFAD is used to gate the id201 detector.}
\label{fig:exp-schem}
\end{figure}

We experimentally characterize the EPMF for a $\beta$-Barium Borate (BBO) crystal cut for frequency-degenerate non-collinear type-II phase matching at $1550$ nm. The experimental setup is shown in Fig.~\ref{fig:exp-schem}. The $5$ mm long crystal cut at $29.67$\textdegree\ was pumped with a $60$ mW (CW) tunable, fiber-coupled laser (Coherent, Mira900) with a spatial mode radius of $150$ $\mu$m.  Collection optics and fibers (Corning, SMF-28e) allowed the collection of  photons emerging at an angle of $3$\textdegree\ with respect to the pump propagation direction, in a spatial mode with a radius of $105$ $\mu$m. 

Although various techniques for joint spectrum measurements are well known, the infrared  range has been inaccessible due to the immaturity of detection techniques, causing high dark count rates and low efficiency. Here we use a free-running negative feedback avalanche diode (NFAD) \cite{Yan2012}  in combination with a gated id201 detector. This configuration is optimal for the signal to noise ratio due to the NFAD's low dark count rate ($100$/sec) and the id201's high quantum efficiency ($15\%$).

The EPMF diagonal slices $\Theta(\omega_s,\omega_p-\omega_s)$ can be accessed by measuring the joint spectral amplitudes for a range of CW pump laser settings.
For that we measure the relative detection time statistics of photons traveling trough two long fibers and exploit the dispersion experienced in single mode fiber \cite{Avenhaus2009}. 


When CW pumping is used, this spectrometer needs to be calibrated using a known, pulsed laser.
Moreover, it is crucial to know the dependence of the group velocity on the wavelength in the fibers, which we characterize by solving the Helmholtz equation. 
Then by using two pulsed lasers at $1310$ nm and $1550$ nm it was possible to experimentally measure two group velocities. These results agreed with the theoretical prediction within $0.7\%$. 
The spectral resolution is estimated to be $4.1$ nm, determined by 
the joint effect of dispersion, the lengths of the fibers ($4202$ m and $4217$ m), the jitter of the detectors (NFAD~$<156$ ps and id201~$<300$ ps) and the time resolution of the time tagging unit ($156$ ps). This can in principle be improved to $0.86$ nm using a $20$ km long fiber.

\begin{figure}
\includegraphics[width=0.9\columnwidth]{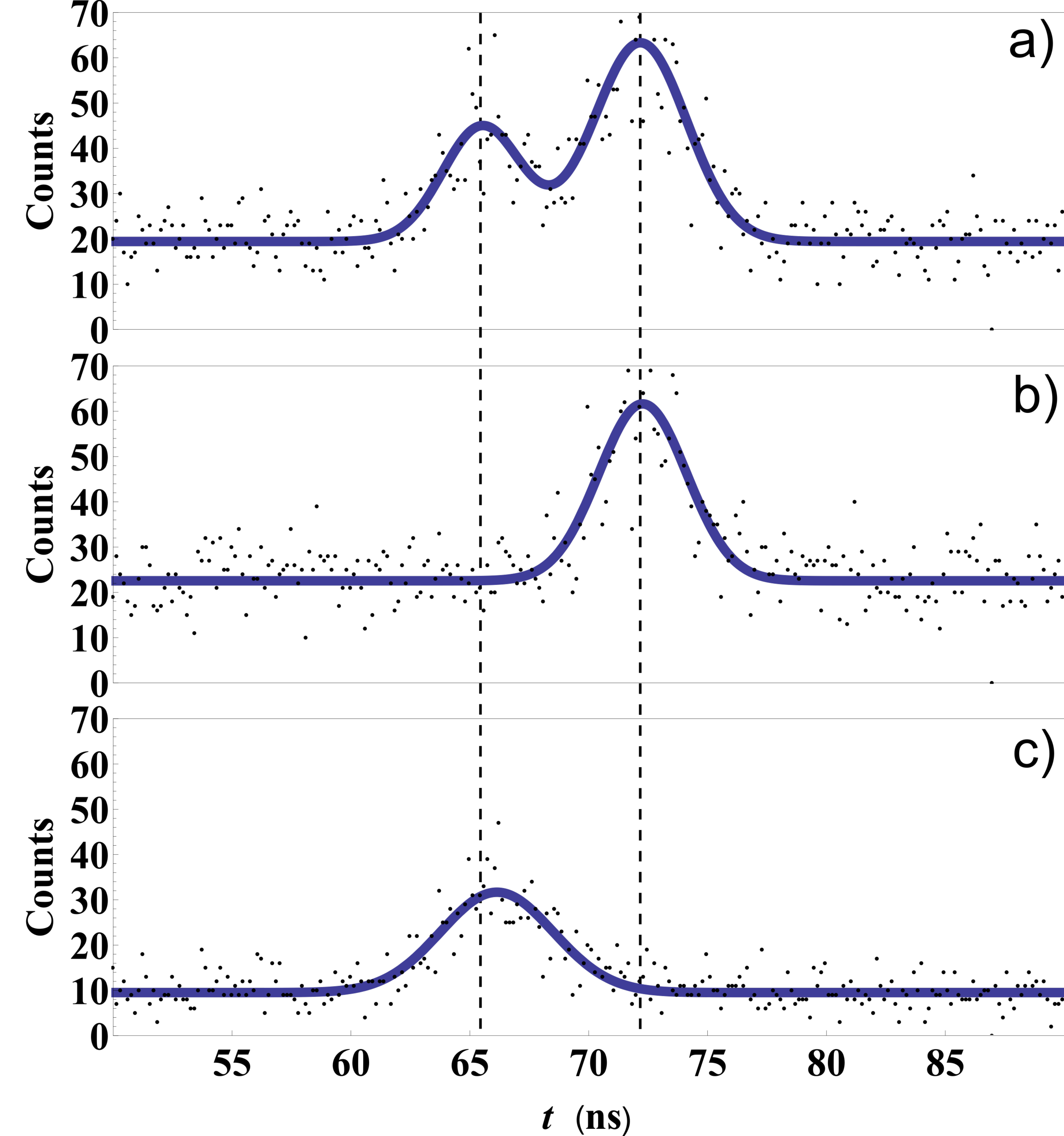}
\caption{(Color online) Histograms (the dots are measured data, poissonian error bars are omitted due to visibility, the lines are gaussian fits to the data) of photon arrival times with 200 s acquisition time and 156 ps bin size. a) Triggering with either e- or o-polarized photon. Placing the polarizer in front of triggering coupler allows the selection of b) e-polarized c) o-polarized photons.}
\label{plot:polarizers}
\end{figure}

Before we characterized the EPMF of the source, we did two tests to confirm that the photons were collected from the intersection of the SPDC cones. The crystal was tilted such that extraordinary (e) and ordinary (o) polarized photons were frequency non-degenerate, with wavelengths $1538$ nm (o) and $1561$ nm (e). Due to the different propagation velocities of extraordinary and ordinary photons in the fiber, two  peaks are visible in the timing histogram, plotted in Fig.~\ref{plot:polarizers} (a). The e-polarized (o-polarized) photons were then selected as a trigger by putting a vertical (horizontal) polarizer in front of the coupler connected to the NFAD, which can be seen in Fig.~\ref{plot:polarizers} (b,c). One peak of the histogram vanished depending on the polarizer's orientation. It is clearly visible that histograms (b) and (c) add up to the one presented in (a). Note that the two peaks are different in height, which can be attributed to a walk off effect in the BBO crystal.

\begin{figure}[h!]
\centering
\includegraphics[width=1\columnwidth]{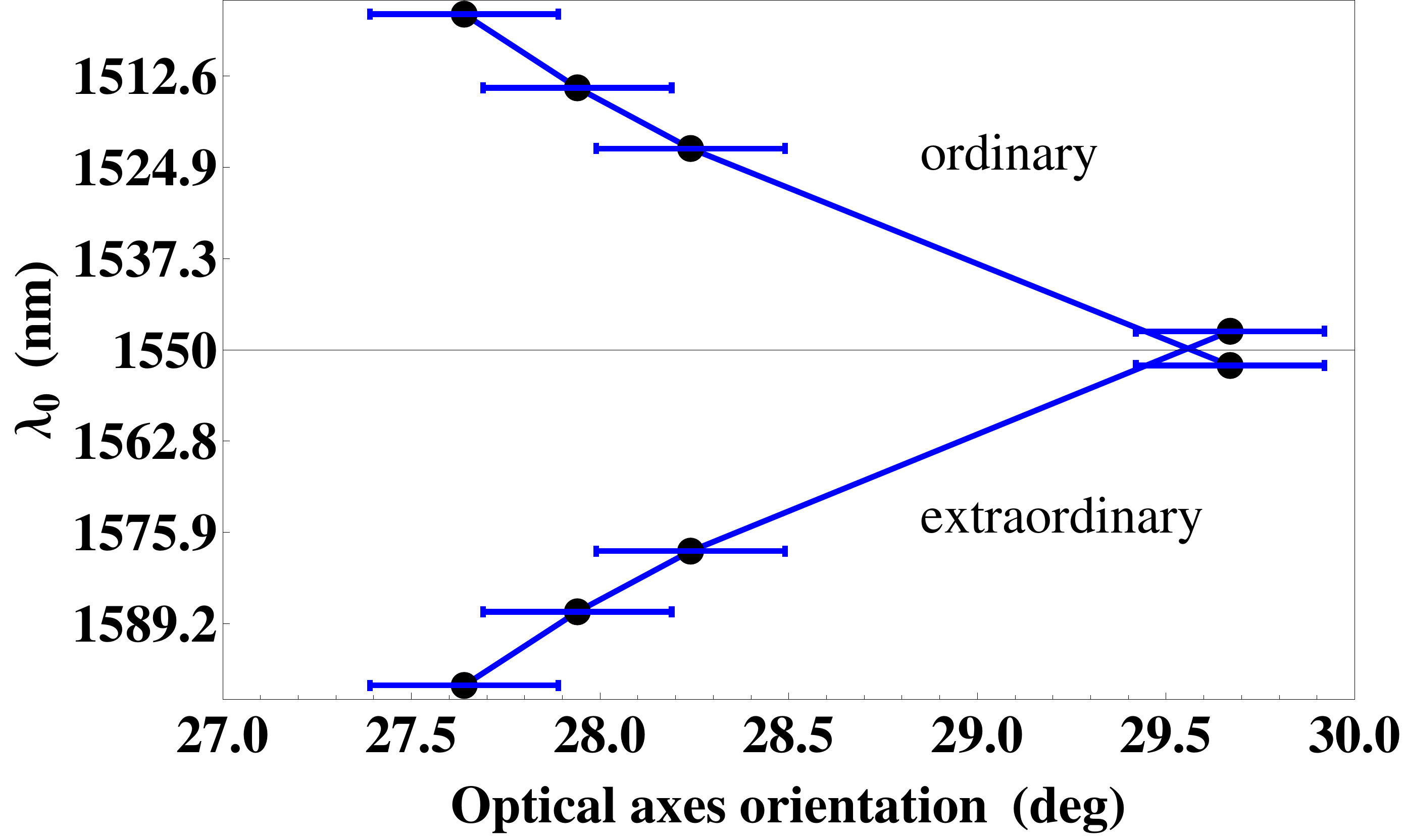}
\caption{(Color online) Measured central frequencies (black cirlces) and guiding blue line of extraordinary and ordinary photons depending on the internal angle between pump beam and optical axes.  The wavelength error is smaller than the marker size. }
\label{plot:crystchangewl}
\end{figure}

After selecting the e-polarized photon for a trigger, we tilted the crystal to achieve spectrally degenerate emission at a pump wavelength of $775$ nm. The measurement results are depicted in Fig.~\ref{plot:crystchangewl}.
The internal angle between pump beam and the optical axes was measured using back reflection with an accuracy of $0.25$\textdegree. The central frequencies clearly moved together as the crystal was tilted towards the degeneracy point. This shows the tuning capabilities of the BBO crystal in the infrared range.

Once the crystal was set for the degenerate emission, the EPMF was characterized by several joint spectrum measurements for the CW pump in the range of $687-796$ nm. The lower pump wavelength limit is given by the tuning capabilities of the Ti:Sapphire laser, whereas the upper limit is set by the quantum efficiency of the \mbox{InGaAs} detectors.
\begin{figure}
\includegraphics[width=1\columnwidth]{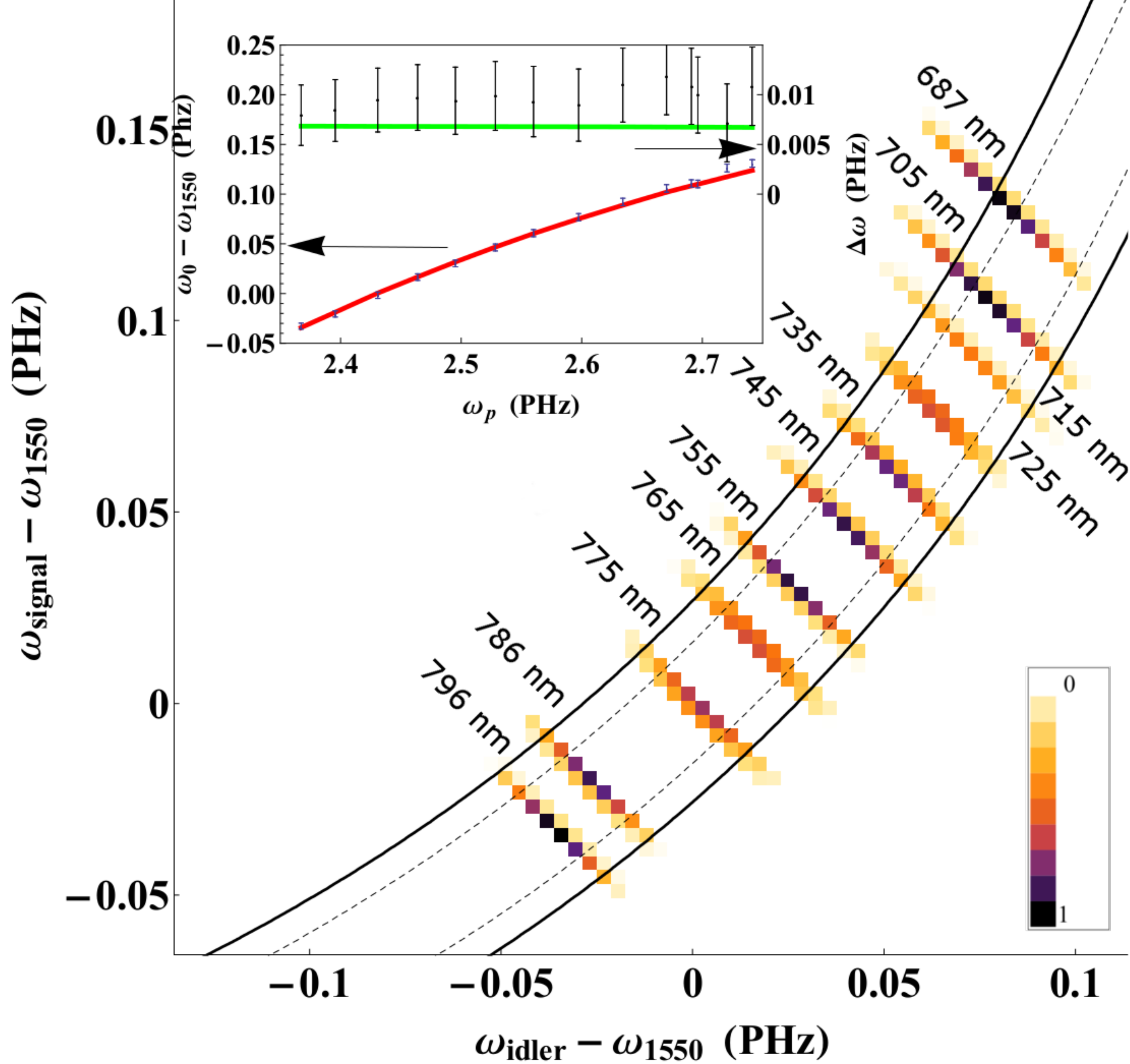}
\caption{(Color online) Effective phase matching function measurement, (colored squares) and theory (continuous contour at 13.5\% dashed contour line at 50\%). The cuts labelled with pump wavelength $\lambda_p$). The sizes of the squares represent the resolution of the spectrometer. The inset shows the comparison between measured  and theoretical central frequencies and EPMF widths. This particular shape of the EPM curve allows for the generation of positively spectrally correlated photons.}
\label{fig:epm-comp}
\end{figure}
During the measurements, the NFAD registered around $15,000$ cps, which were used to gate the id201, which detected $60$ cps. Taking into account detection efficiencies, this corresponds to roughly $1,000,000$ photon pairs produced per second. The expected fiber coupling efficiency is $3 \%$,  which is reduced by experimental imperfections to $0.8\%$ observed in experiment. 
The measurement results are depicted in Fig.~\ref{fig:epm-comp} and agree well with the theoretical model ~\cite{Kolenderski2009,Kolenderski2009a,Kolenderski}. The data acquisition time for each EPMF slice was about $6$ minutes. The measured widths of the EPMF exceed the theoretical widths due to the resolution of the spectrometer. However this effect does not affect its central position. This leads to a very good agreement between measured and theoretical central frequencies. The EPMF in this configuration allows in particular for the positive spectral correlations or decorrelation  within photon pairs without using spectral filtering. 

In summary, we experimentally demonstrated a phase matching condition in type-II BBO at $1550$ nm which allows for the generation of fiber-coupled photons pairs featuring positive spectral correlations. This measurement was only possible because of the improved sensitivity of our fiber spectrometer, which was achieved by using a combination of free-running and gated \mbox{InGaAs} detectors. Using a femtosecond pump laser, positively spectrally correlated or decorrelated photon pairs can be generated. 

\smallskip
The authors  acknowledge  funding from NSERC (CGS, QuantumWorks, Discovery, USRA), Ontario Ministry of Research and Innovation (ERA program and research infrastructure program), CIFAR, Industry Canada and the CFI.  PK acknowledges support by Mobility Plus project financed by Polish Ministry of Science and Higher Education. We also thank Zhizhong Yan and Deny Hamel for the fruitful discussions about the NFADs.

\bibliographystyle{ol}	
\bibliography{SpectCorrTelecom}

\end{document}